\documentclass[letterpaper,11pt]{article}
\usepackage{jheppub}
\usepackage{array}
\usepackage{amssymb}
\usepackage{amsmath}
\usepackage{amsfonts}
\usepackage{graphics}
\usepackage{graphicx}
\usepackage{bm}
\usepackage{bbm}

\makeatletter
\def\@fpheader{\relax}
\makeatother

\title{Electroweak phase transition in a model with gauged lepton number}
\author[a,b]{Alfredo Aranda,}
\author[c]{Enrique Jim\'enez,}
\author[a]{Carlos A. Vaquera-Araujo}
\affiliation[a] {Facultad de Ciencias, CUICBAS, Universidad de Colima\\
Bernal D\'iaz del Castillo 340, Colima 28045, M\'exico }
\affiliation[b] {Dual C-P Institute of High Energy Physics, Colima 28045, M\'exico }
\affiliation[c] {Instituto de F\'isica, Universidad Nacional Aut\'onoma de M\'exico,\\ Apdo. Postal 20-364, 01000, M\'exico D.F., M\'exico}

\emailAdd{fefo@ucol.mx, ejimenez@fisica.unam.mx, cvaquera@ucol.mx}

\abstract{In this work we study the electroweak phase transition in a model with gauged lepton number. Here, a family of vector-like leptons is required in order to cancel the gauge anomalies. Furthermore, these leptons can play an important role in the transition process. We find that this framework is able to provide a strong transition, but only for a very limited number of cases. }

\begin{document}
\maketitle
\flushbottom

\section{Introduction}

An important hint in the search for new physics is the observation that the Standard Model (SM) does not fulfill the requirement of thermal equilibrium departure, needed for the dynamical generation of the baryon asymmetry of the universe \cite{Sakharov:1967dj}. In the context of electroweak baryogenesis, such departure requires the existence of a strong first order electroweak phase transition (EWPT), and the SM is unable to undergo a transition of that kind \cite{Kajantie:1996mn,Csikor:1998eu}.

Several works have been devoted to explore the possibility of accommodating a strong first order EWPT in SM extensions, and in particular, it has been shown in  \cite{Davoudiasl:2012tu, Carena:2004ha}  that new fermionic degrees of freedom can accomplish that task. In this regard, an interesting option is the inclusion of vector-like fermions, i.e. fermions with chiral components transforming equally under the SM gauge group. A recent study of successful baryogenesis with vector-like fermions can be found in \cite{Fairbairn:2013xaa}, where an additional family of vector-like leptons provides a strong EWPT. That model can be obtained as the low energy limit of a theory of gauged lepton number, defined recently in \cite{FileviezPerez:2011pt,Duerr:2013dza} and further explored in \cite{Schwaller:2013hqa}.    

The realization of the $SU(3)_c\times SU(2)_W\times U(1)_Y\times U(1)_L$ gauge symmetry in \cite{FileviezPerez:2011pt,Duerr:2013dza,Schwaller:2013hqa} is particularly attractive as it contains a dark matter (DM) candidate that emerges naturally (as part of the new lepton sector required to cancel gauge anomalies), its stability being ensured by a remnant discrete symmetry after the spontaneous breaking of $U(1)_L$. 

The purpose of this work is to find out if a strong EWPT can be achieved in the explicit setup of gauged lepton number described in \cite{Schwaller:2013hqa}. Beside SM fields, this minimal model incorporates four exotic leptons (two charged and two neutral), an extra scalar field (that can also contribute to strengthen the transition) and a new neutral massive gauge boson. According to our analysis, a strong EWPT is possible in this model for a very restricted region of the parameter space.

The paper is structured as follows: section \ref{secmod} contains a description of the constituent fields of the model and their available experimental bounds. Section \ref{secV} is dedicated to the construction of the effective potential at finite temperature, approximated at one-loop order, and improved by the thermal correction of scalar masses through the resummation of daisy diagrams. In section \ref{secPT}, the procedure followed in the search of a strong EWPT is outlined, and the obtained results are shown. Finally, our conclusions are presented in section \ref{secCon} and complementary information, regarding field dependent squared masses, is included in appendix \ref{apA}. 

\section{Model}\label{secmod}
In this section, the basic ingredients of the model are defined and their essential features are discussed.

The present work is based on the $SU(3)_c\times SU(2)_W\times U(1)_Y\times U(1)_L$ symmetric model studied in \cite{Schwaller:2013hqa}.  In this minimal model, a SM singlet scalar field with  $(SU(2)_W, U(1)_Y, U(1)_L)$ quantum numbers 
\begin{equation}
\Phi\sim(1,0,3)\, ,
\end{equation}
is responsible for the spontaneous breaking of lepton number gauge symmetry. Anomaly cancellation  \cite{FileviezPerez:2011pt,Duerr:2013dza} is achieved through the introduction of 3 right-handed neutrinos $\nu_{R\, l}\sim  (1,0,1)$ $(l=e,\mu,\tau)$, and a vector-like family of exotic leptons with the following quantum numbers:
\begin{eqnarray}
\ell'_L &\equiv & \left(\nu'_L,\, e'_L \right)^T\sim (2,-1/2,L')\,,\nonumber\\
e'_R &\sim & (1,-1,L'),\qquad \nu'_R \sim (1,0,L')\,, \nonumber\\
\ell''_R &\equiv & \left(\nu''_R,\, e''_R \right)^T\sim (2,-1/2,L'')\,,\nonumber\\
e''_L &\sim & (1,-1,L''),\qquad \nu''_L \sim (1,0,L'')\,,
\end{eqnarray}
where $L'$ and $L''$ are constrained by the anomaly-free condition $L'-L''=-3$.

\subsection{Scalar sector}
Denoting $H\sim(2,1/2,0)$ as the SM Higgs doublet, the tree-level scalar potential is given by
\begin{equation}\label{Pot0}
V^0(H,\Phi)= -\mu_1^2 H^{\dagger}H-\mu_2^2\Phi^{*}\Phi+\lambda_1  (H^{\dagger}H)^2+\lambda_2(\Phi^{*}\Phi)^2+\lambda_3 H^{\dagger}H \Phi^{*}\Phi\,,
\end{equation}
and is positive definite if $\lambda_1\lambda_2>\lambda_3^2/4$.  The scalar fields can be written as
\begin{equation}
H=\left(\begin{array}{c}
G^+\\
\frac{\phi_1+i G_1^0}{\sqrt{2}}
\end{array}\right),\qquad \Phi=\frac{\phi_2+i G_2^0}{\sqrt{2}}\,,
\end{equation}
and their real components obtain vacuum expectation values (VEV's) at zero temperature $\left\langle H\right\rangle=(0,v/\sqrt{2})^T$ and $\left\langle\Phi\right\rangle=u/\sqrt{2}$, determined by the minimum conditions
\begin{equation}\label{MinCond}
\left.\frac{\partial V^0}{\partial\phi_1}\right|_{\begin{smallmatrix}
H=\left\langle H\right\rangle\\\Phi=\left\langle\Phi\right\rangle
\end{smallmatrix}}=\left.\frac{\partial V^0}{\partial\phi_2}\right|_{\begin{smallmatrix}
H=\left\langle H\right\rangle\\\Phi=\left\langle\Phi\right\rangle
\end{smallmatrix}}=0\,,
\end{equation}
with $v=246$ GeV as the electroweak symmetry breaking scale. The above equation can be used to eliminate $\mu_1^2$ and $\mu_2^2$ in favor of the remaining parameters:  
\begin{equation}\label{mu12}
\begin{split}
\mu_1^2&=\frac{\lambda_3}{2} u^2+\lambda_1 v^2\,,\\
\mu_2^2&=\frac{\lambda_3}{2} v^2+\lambda_2  u^2\,.
\end{split}
\end{equation}
Substituting these relations in (\ref{Pot0}), up to a constant term, the tree level potential for the real scalar fields $\phi_1$ and $\phi_2$ becomes
\begin{equation}\label{Vtree}
V_{\text{tree}}(\phi_1,\phi_2)=\frac{\lambda_1 }{4}\left(\phi_1^2-v^2\right)^2+ \frac{\lambda_2}{4}\left(\phi_2^2-u^2\right)^2+\frac{\lambda_3}{4}\left(\phi_1^2-v^2\right)\left(\phi_2^2-u^2\right)\,.
\end{equation}

There are two neutral CP-even scalars $h_1$ and $h_2$. From eq.(\ref{Vtree}), their tree-level mixing and masses are given by
\begin{equation}
\left(
\begin{array}{c}
h_1\\
h_2
\end{array}
\right)=
\left(
\begin{array}{cc}
\cos\theta &-\sin\theta \\
\sin\theta & \cos\theta
\end{array}
\right)\left(
\begin{array}{c}
\phi^0_1\\
\phi^0_2
\end{array}
\right)\,, \qquad \tan2\theta=\frac{\lambda_3 v u}{\lambda_2 u^2-\lambda_1 v^2}\,,
\end{equation}
\begin{equation}\label{scalmasstree}
m^2_{h_1,h_2}=\lambda_1 v^2+\lambda_2  u^2\mp\sqrt{\left(\lambda_2  u^2-\lambda_1 v^2\right)^2+ \lambda_3^2 u^2 v^2}\,,
\end{equation}
with $\phi^0_1\equiv\phi_1-v$ and $\phi^0_2\equiv\phi_2-u$ as the gauge eigenstates. We assume that the lightest mass eigenstate $h_1$ is mostly composed of $\phi^0_1$, and we identify it with the spin 0 resonance discovered at the LHC  \cite{Aad:2012tfa,Chatrchyan:2012ufa}, fixing its mass at $m_{h_1}=125.7$ GeV \cite{Agashe:2014kda}.

\subsection{Gauge sector}

The interactions of $Z_L$, the gauge boson associated to $U(1)_L$, are encoded in the Lagrangian density
\begin{equation}
\mathcal{L}_{Z_L}=(D^\mu\Phi)^\dagger (D_\mu\Phi)+\frac{\epsilon}{2}Z_L^{\mu\nu}B_{\mu\nu}+\bar{\ell}_L	\gamma^\mu D_\mu \ell_L	+\bar{\ell}'_L	\gamma^\mu D_\mu \ell'_L+\bar{\ell}''_R\gamma^\mu D_\mu \ell''_R	\, .												
\end{equation}  
Here, the covariant derivative is defined as $D^\mu=\partial^\mu+ig_W \frac{\tau^a}{2} W^{a\mu}+ig_Y Y B^\mu+i g_L LZ_L^\mu$. The $U(1)_L$ and $U(1)_Y$  field strength tensors are  denoted by $Z_L^{\mu\nu}$ and $B^{\mu\nu}$, respectively. Note that left handed SM  lepton  doublets $\ell_L$  couple to $Z_L$, and that because of the $U(1)$ kinetic mixing, there is a $B^\mu -Z_L^\mu$ coupling parameterized by $\epsilon$. 

As observed in \cite{Schwaller:2013hqa}, the strongest bound on $Z_L$ come from LEP II data \cite{Carena:2004xs}:
\begin{equation}
u\geq 1.7 \,\text{TeV}.
\end{equation}
The above condition is roughly independent of the $g_L$ coupling value. Direct searches for dark matter and precision measurements of the $Z$-pole constrain the kinetic mixing to be small. Typical values consistent with the current bounds are  $\epsilon\lesssim 10^{-2}$  \cite{Schwaller:2013hqa,Cline:2014dwa}.

In the following, we take $\epsilon=0$ at tree level, and set the scale of $U(1)_L$ symmetry breaking at its lower bound of $u= 1.7 \,\text{TeV}$. In this limit, the mass of $Z_L$ turns out to be independent of the electroweak scale:
\begin{equation}
m_{Z_L}^2=9g_L^2u^2,
\end{equation} 
and consequently, the role played by $Z_L$ in the EWPT becomes negligible. For completeness, in the numerical analysis performed in this work, we include the $Z_L$ contribution to the one-loop effective potential with $g_L=0.4$ .

\subsection{Vector-like fermion sector}

The Lagrangian density that describes the Yukawa interactions of the vector-like family of leptons is
\begin{equation}\label{LYuk}
\begin{split}
\mathcal{L}_{\cal{Y}}=&-y'_e\overline{\ell'_L} H e'_R-y''_e\overline{\ell''_R} H e''_L-y'_\nu\overline{\ell'_L }\widetilde{H} \nu'_R-y''_\nu\overline{\ell''_R} \widetilde{H} \nu''_L	\\
&-c_\ell \overline{\ell'_L}\Phi\ell''_R-c_e \overline{e''_L} \Phi e'_R-c_ \nu  \overline{\nu''_L} \Phi \nu'_R+ \text{h.c.}\, ,
\end{split}										
\end{equation}  
with $\widetilde{H}\equiv i\sigma^2H^*$. Additional terms can be introduced in $\mathcal{L}_{\cal{Y}}$ upon a definite choice of $L'$ and $L''$. A discussion about these terms can be found in \cite{Schwaller:2013hqa}. In this work we leave $L'$ and $L''$ unspecified beyond the anomaly-free condition $L'-L''=-3$, and therefore our analysis is restricted to this minimal set of terms. For simplicity, we assume that all the Yukawa couplings of the vector-like fermion sector are real.

From eq.(\ref{LYuk}), the mass matrices for the new neutral and charged leptons are given by
\begin{equation}
\left(\overline{\nu'_L}\,\,\overline{\nu''_L}\right)\mathcal{M}_n\left(\begin{array}{c}\nu'_R \\ \nu''_R\end{array}\right)+\text{h.c.}\, ,\qquad
\mathcal{M}_n=\frac{1}{\sqrt{2}}
\left(\begin{array}{cc}
y'_\nu v  & c_\ell u\\
c_\nu u  & y''_\nu v
\end{array}\right)\, ,
\end{equation}

\begin{equation}
\left(\overline{e'_L}\,\,\overline{e''_L}\right)\mathcal{M}_e\left(\begin{array}{c}e'_R \\ e''_R\end{array}\right)+\text{h.c.}\, ,\qquad
\mathcal{M}_e=\frac{1}{\sqrt{2}}
\left(\begin{array}{cc}
y'_e v  & c_\ell u\\
c_e u & y''_e v
\end{array}\right)\, .
\end{equation}
The eigenvalues of $\mathcal{M}_n$ ($\mathcal{M}_e$) can be obtained through the singular value decomposition $\mathcal{M}_N=U^{n\dagger}_L \mathcal{M}_n U^n_R$ ($\mathcal{M}_E=U^{e\dagger}_L \mathcal{M}_e U^e_R$), where $U^n_L$, $U^n_R$ ($U^e_L$, $U^e_R$) are independent unitary matrices and $\mathcal{M}_N$ ($\mathcal{M}_E$) is a diagonal matrix with positive entries $m_{N_{1,2}}$ ($m_{E_{1,2}}$). In the present study, we adopt the following constraints for the physical masses of the heavy leptons: 
\begin{equation}\label{vlfbounds}
m_{N_{1,2}}>\frac{M_Z}{2}\, ,\qquad
m_{E_{1,2}}>100\, \text{GeV}\, .
\end{equation}

After spontaneous symmetry breaking, a residual $Z_2$ symmetry is preserved in the model. All SM fields are even under this discrete symmetry, while vector-like leptons are odd. This feature guarantees that the lightest state of the new lepton sector is stable. We identify such stable lepton with the neutral state $N_1$, rendering it appropriate as a DM candidate. 

\section{Finite temperature one-loop potential}\label{secV}

In order to study the EWPT, we use the scalar potential at finite
temperature, approximated at one-loop and including daisy resummations for the scalar fields.
The one-loop zero temperature corrections to the scalar potential have the form
\begin{equation}\label{1loop}
V_{\text{1-loop}}=\sum_{i}\frac{n_i}{64\pi^2}m_i(\phi_1,\phi_2)^4\left[\log\frac{m_i(\phi_1,\phi_2)^2}{\Lambda^2}-\frac{3}{2}\right]\, ,
\end{equation}
where $i$ labels the contributing particles, $n_i$ stands for their number of degrees of freedom (including a $\pm$ sign due to statistics), and $m_i(\phi_1,\phi_2)$ denotes their so-called field dependent masses. The scale is chosen as $\Lambda=6 $ TeV in our analysis \footnote{Here $\Lambda$ is an arbitrary scale. The true renormalization scale is introduced by the conditions eq. (\ref{rencond}). We set $\Lambda$ as a typical scale greater than the highest mass involved. The final results are largely insensitive to this choice, as $\Lambda$ contributes only through a constant term in the renormalized one-loop zero temperature correction to the scalar potential $V_{\text{1-loop}}+V_{\text{CT}}$. }. Among the SM fields we take into account the gauge bosons $W^\pm$ and $Z$ with $n_W=6$ and $n_Z=3$, respectively, as well as the top quark $t$ with $n_t=-12$. The field dependent squared masses for scalars and vector-like leptons are listed in appendix \ref{apA}.

The one-loop potential in eq.(\ref{1loop}) is complemented with counterterms 
\begin{equation}\label{CT}
V_{\text{CT}}=\frac{1}{2}\alpha_1\phi_1^2+\frac{1}{2}\alpha_2\phi_2^2+\frac{1}{4}\beta_1\phi_1^4+\frac{1}{4}\beta_2\phi_2^4+\frac{1}{4}\beta_3\phi_1^2\phi_2^2\, ,
\end{equation}
that preserve the location of the minimum and the masses of the scalars:
\begin{equation}\label{rencond}
\left.\frac{\partial}{\partial \phi_a}\left(V_{\text{1-loop}}+V_{\text{CT}}\right)\right|_{\begin{smallmatrix}
\phi_1=v\\\phi_2=u
\end{smallmatrix}}=0\, ,
\end{equation}
\begin{equation}
\left.\frac{\partial^2}{\partial \phi_{a}\partial \phi_{b}}\left(V_{\text{1-loop}}+V_{\text{CT}}\right)\right|_{\begin{smallmatrix}
\phi_1=v\\\phi_2=u
\end{smallmatrix}}=0\, , \qquad a,b=1,2\, .
\end{equation}
Thus, the coefficients in $V_{\text{CT}}$ are
\begin{equation}
\begin{split}
\alpha_1&=\frac{1}{2v}\left[v\omega^{(2,0)}+u\omega^{(1,1)}-3\omega^{(1,0)}\right]\, ,\\
\alpha_2&=\frac{1}{2u}\left[v\omega^{(1,1)}+u\omega^{(0,2)}-3\omega^{(0,1)}\right]\, ,\\
\beta_1&=\frac{1}{2v^3}\left[\omega^{(1,0)}-v\omega^{(2,0)}\right]\, ,\\
\beta_2&=\frac{1}{2u^3}\left[\omega^{(0,1)}-u\omega^{(0,2)}\right]\, ,\\
\beta_3&=-\frac{\omega^{(1,1)}}{vu}\, ,
\end{split}
\end{equation}
with
\begin{equation}
\omega^{(j,k)}=\left.\frac{\partial^{j+k} }{\partial\phi_1^j\partial\phi_2^k}V_{\text{1-loop}}\right|_{\begin{smallmatrix}
\phi_1=v\\\phi_2=u
\end{smallmatrix}}
\, . 
\end{equation}
Following the prescription of \cite{Cline:2011mm}, an infrared cutoff $m^2_{\text{IR}}=m^2_{h_1}$ is imposed in the determination of the counterterms related to Goldstone boson contributions.

The one-loop finite temperature corrections to the effective potential can be written as
\begin{equation}
V_{T}=\frac{T^4}{2\pi^2 }\left[\sum_{i} n_i I_{-} \left(\frac{m^2_i(\phi_1,\phi_2)}{T^2}\right)+\sum_{j} n_j I_{+} \left(\frac{m^2_j(\phi_1,\phi_2)}{T^2}\right) \right]\,,
\end{equation}
where $i$ ($j$) runs over all boson (fermion) fields. Thermal corrections are dictated by the integral $I_{-}$ ($I_{+}$), defined as 
\begin{equation}\label{Tint}
I_{\mp}(\alpha)=\int_0^\infty x^2\log\left(1\mp e^{-\sqrt{x^2+\alpha}}\right).
\end{equation}
As the numerical evaluation of this integral is computationally expensive, it is often useful to work with an approximate form. The high temperature ($\alpha\ll 1$) expansion of eq.(\ref{Tint}) is
\begin{equation}
I_{-}^{HT}(\alpha)=-\frac{\pi ^4}{45} +\frac{\pi ^2  }{12}\alpha-\frac{\pi }{6}\alpha ^{\frac{3}{2}}-\frac{1}{32} \alpha ^2 \log \left(\frac{\alpha}{c_B}\right)\, ,
\end{equation}
\begin{equation}
I_{+}^{HT}(\alpha)=\frac{7 \pi ^4}{360}-\frac{\pi ^2  }{24}\alpha-\frac{1}{32} \alpha ^2 \log \left(\frac{\alpha}{c_F} \right)\, , 
\end{equation}
with $c_F=\pi^2\exp\left(3/2-2\gamma\right)$ and $c_B=16c_F$. At low temperature ($\alpha\gg 1$), the same integral can be approximated as
\begin{equation}
I_{+}^{LT}(\alpha)=-I_{-}^{LT}(\alpha)= \sqrt{\frac{\pi }{2}} e^{-\sqrt{\alpha }} \alpha ^{3/4} \left(1+\frac{15}{8 \sqrt{\alpha }}+\frac{105}{128 \alpha }\right)\, .
\end{equation}
For the numerical analysis, we work with the following smooth interpolation between the two regimes \cite{Li:2014wia}:
\begin{equation}
I_{\mp}^{\text{app}}(\alpha)=t_{\mp}(\alpha)I_{\mp}^{HT}(\alpha)+[1-t_{\mp}(\alpha)]I_{\mp}^{LT}(\alpha)\, ,
\end{equation}
where $t_{-}=e^{-(\alpha/6.3)^4}$ and $t_{+}=e^{-(\alpha/3.25)^4}$. This approximation deviates from the exact integral by no more than  4\% in the worst case, and for most values the difference is much smaller.

As described in \citep{Cline:2011mm}, the one-loop potential at finite temperature can be further improved through the inclusion of thermal corrections to the scalar masses, coming from the resummation of daisy diagrams, and fermions do not acquire thermal mass corrections. The corrections are included in the field dependent masses listed in appendix \ref{apA}. These thermally corrected masses are the ones that contribute to $V_{\text{1-loop}}$ and $V_{T}$. The final effective potential is given by
\begin{equation}
V(\phi_1,\phi_2,T)=V_{\text{tree}}+V_{\text{1-loop}}+V_{\text{CT}}+V_{T}\, .
\end{equation}

\section{Electroweak phase transition}\label{secPT}

\subsection{Parameter scan}

After fixing the mass of the lightest scalar, there are 9 free parameters in the present scheme: the seven Yukawa couplings $y'_\nu$, $y''_\nu$, $y'_e$, $y''_e$, $c_\ell$, $c_\nu$, $c_e$ and two quartic couplings that can be chosen among $\lambda_1$, $\lambda_2$ and $\lambda_3$. All free parameters are restricted to lie in the perturbative interval $P=(-\sqrt{4\pi},\sqrt{4\pi})$. We eliminate $\lambda_1$ from eq.(\ref{scalmasstree}) in terms of $m^2_{h_1}$, $\lambda_2$ and $\lambda_3$:
\begin{equation}
\lambda_1=\frac{1}{2}\left(\frac{m^2_{h_1}}{v^2}+\frac{\lambda_3^2 u^2}{2\lambda_2u^2-m^2_{h_1}}\right),
\end{equation}
and scan randomly $\lambda_2$ and $\lambda_3$ in the region
\begin{equation}
0<\lambda_2<\sqrt{4\pi}\, ,\qquad 0\leq\lambda_3<\sqrt{4\pi}\, ,
\end{equation}
subject to the following constraints:  $\lambda_2  u^2-\lambda_1 v^2>0$ (as the lightest scalar is mostly $\phi_1^0$), $\lambda_3^2/2<\lambda_1\lambda_2$ (the positive definite condition of the tree-level potential) and $\lambda_1^2<4\pi$.  Notice that negative values for $\lambda_3$ are not allowed once $\lambda_2  u^2-\lambda_1 v^2>0$ is imposed, together with the positivity of the quadratic coefficients $\mu_1^2$ and $\mu_2^2$ in eq.(\ref{mu12}). All Yukawas from the vector-like sector are picked from $P$ in a random fashion, restricted to eq.(\ref{vlfbounds}).

From all the generated points, we first test if the potential is metastable (in the sense of \cite{Dorsch:2013wja}) by searching randomly for a deeper minimum in a region of radius $r=\sqrt{\phi_1^2+\phi_2^2}=\Lambda$. If the test is passed, the point is regarded as physical, otherwise it is discarded.

In this work, we focus exclusively in the electroweak phase transition. The dominant mechanism behind the transition is related to the exotic leptons as described in \cite{Davoudiasl:2012tu} (see \cite{Fairbairn:2013xaa} for a detailed discussion).  We assume that at a sufficiently high temperature $T_h$, where lepton number is already broken, the minimum of the effective potential is located at $V(0,u_1,T_h)$. As the temperature decreases, another minimum arises at $V(v_2,u_2,T_l)$. A first order phase transition occurs if there is a critical temperature $T_c$ such that both minima become degenerate $V(0,u_1,T_c)=V(v_2,u_2,T_c)$, and the distance between them is non zero. If the ratio
\begin{equation}
\xi(T_c)=\frac{v_2}{T_c}
\end{equation}
satisfies $\xi(T_c)>1$, the phase transition is considered to be strong \cite{Ahriche:2007jp}.  

We search for a phase transition by following the minimum of the potential (starting from zero-temperature) in consecutive incremental steps of $T_{\text{step}}=10\, \text{GeV}$, until either the difference between both minima $\Delta V(T)\equiv V(0,u_1,T)-V(v_2,u_2,T)$  becomes negative, or a  high temperature $T_{\text{max}}=300\, \text{GeV}$ is reached, at which a strong first order transition becomes improbable.  If a phase transition is identified, the process is refined with steps of  $T_{\text{step}}=1\, \text{GeV}$, starting from the last point with $\Delta V(T)>0$. Finally, the critical temperature $T_c$ is identified with the highest temperature satisfying $\Delta V(T)>0$, and the ratio $\xi$ is evaluated at $T_c$ and $T_c+1\, \text{GeV}$. If the relation $\xi>1$ is satisfied in both cases, the phase transition is determined to be of strong first order. 

\subsection{Results}
\begin{figure}[h]
	\centering
	    	\includegraphics[scale=0.6]{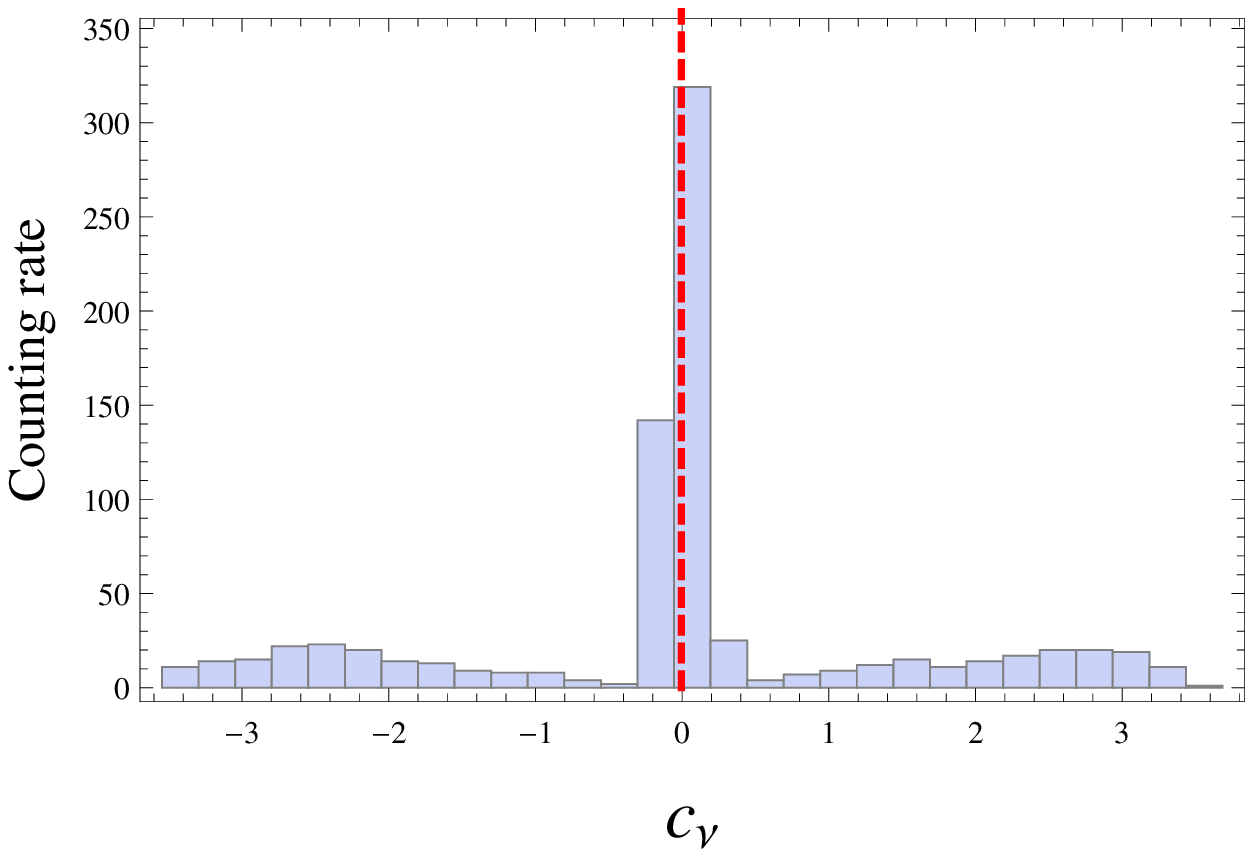} 
		\includegraphics[scale=0.6]{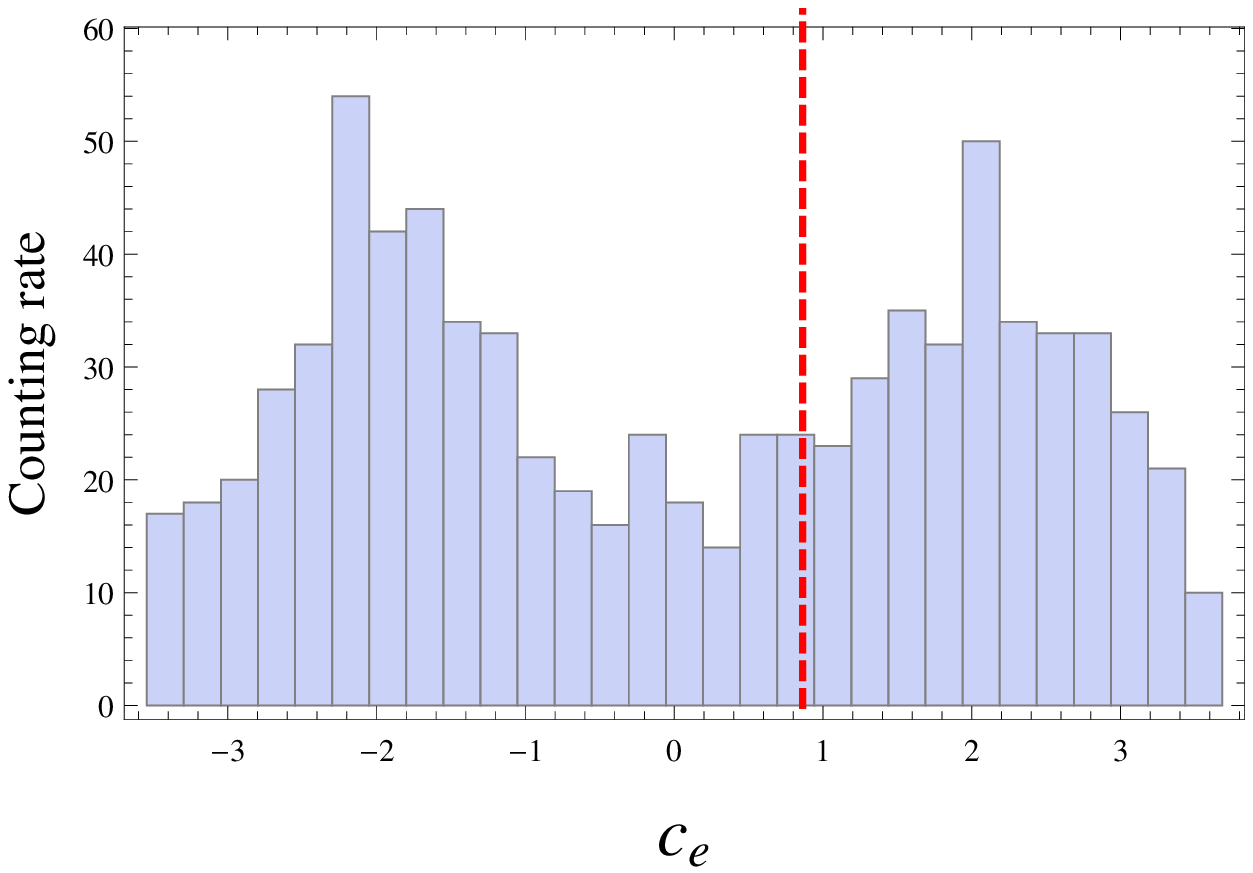} \\
		\includegraphics[scale=0.6]{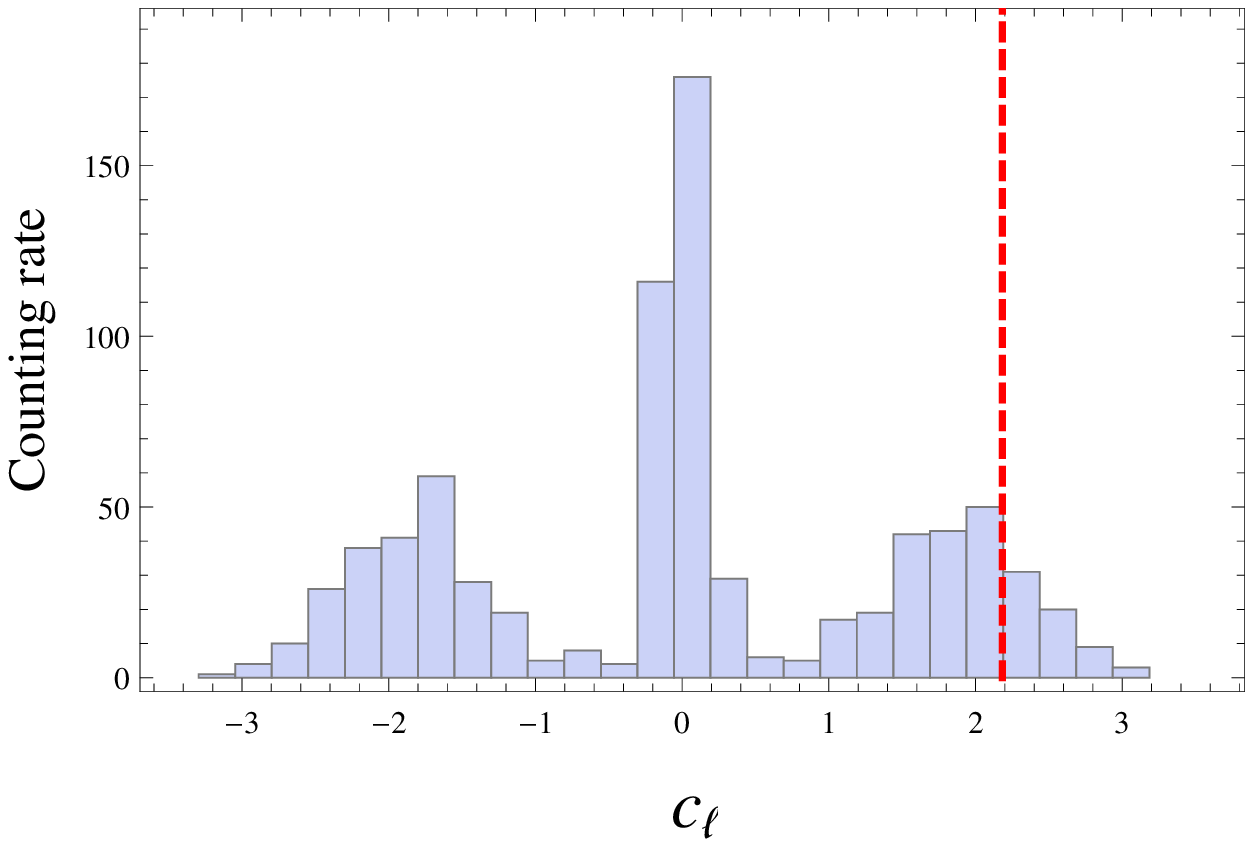} 
	\caption{\small{Counting rates for strong phase transition points as a function 
of the Yukawa couplings  $c_\nu$ (left), $c_e$ (right) and $c_\ell$ (bottom). The couplings of the benchmark point in eq. (\ref{BP}) are indicated as vertical lines.}}
	\label{fig:c}
\end{figure} 

\begin{figure}[h]
	\centering
	    	\includegraphics[scale=0.6]{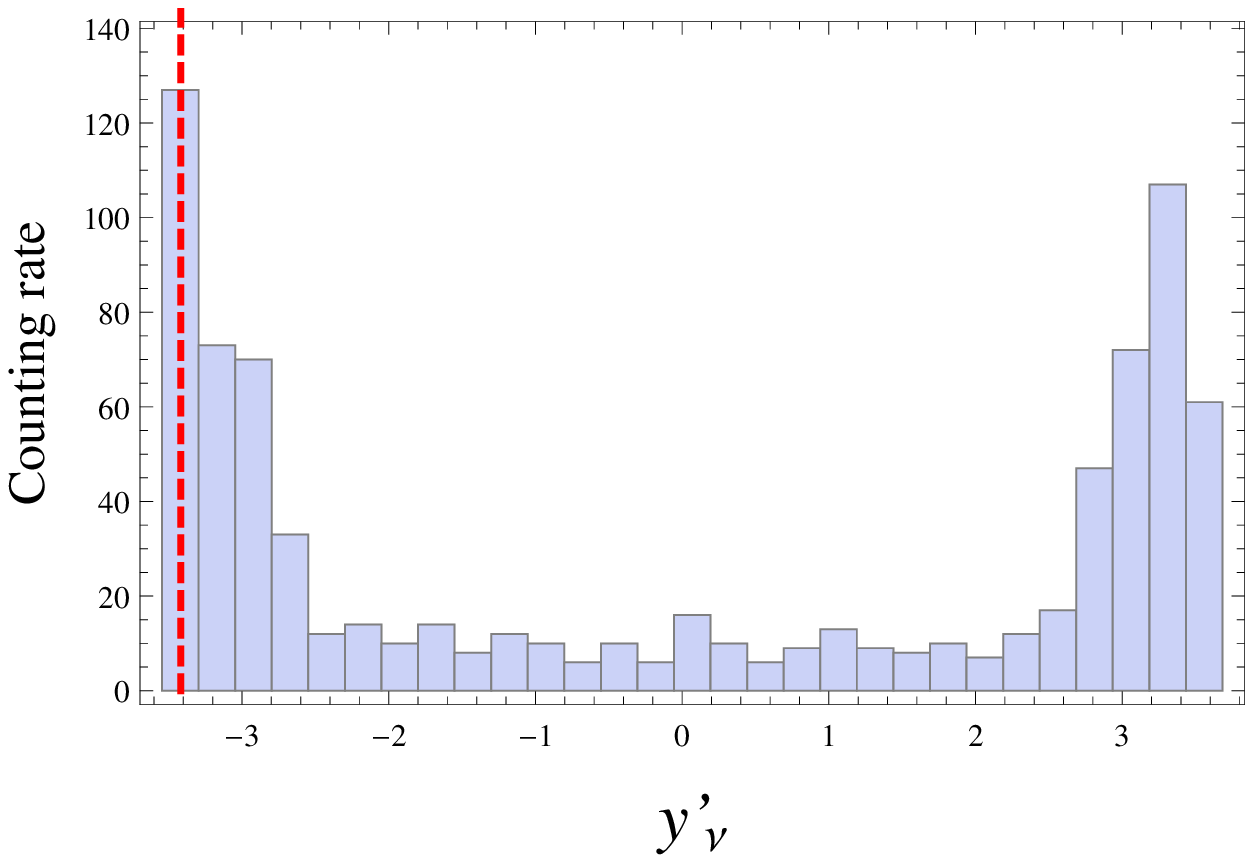} 
		\includegraphics[scale=0.6]{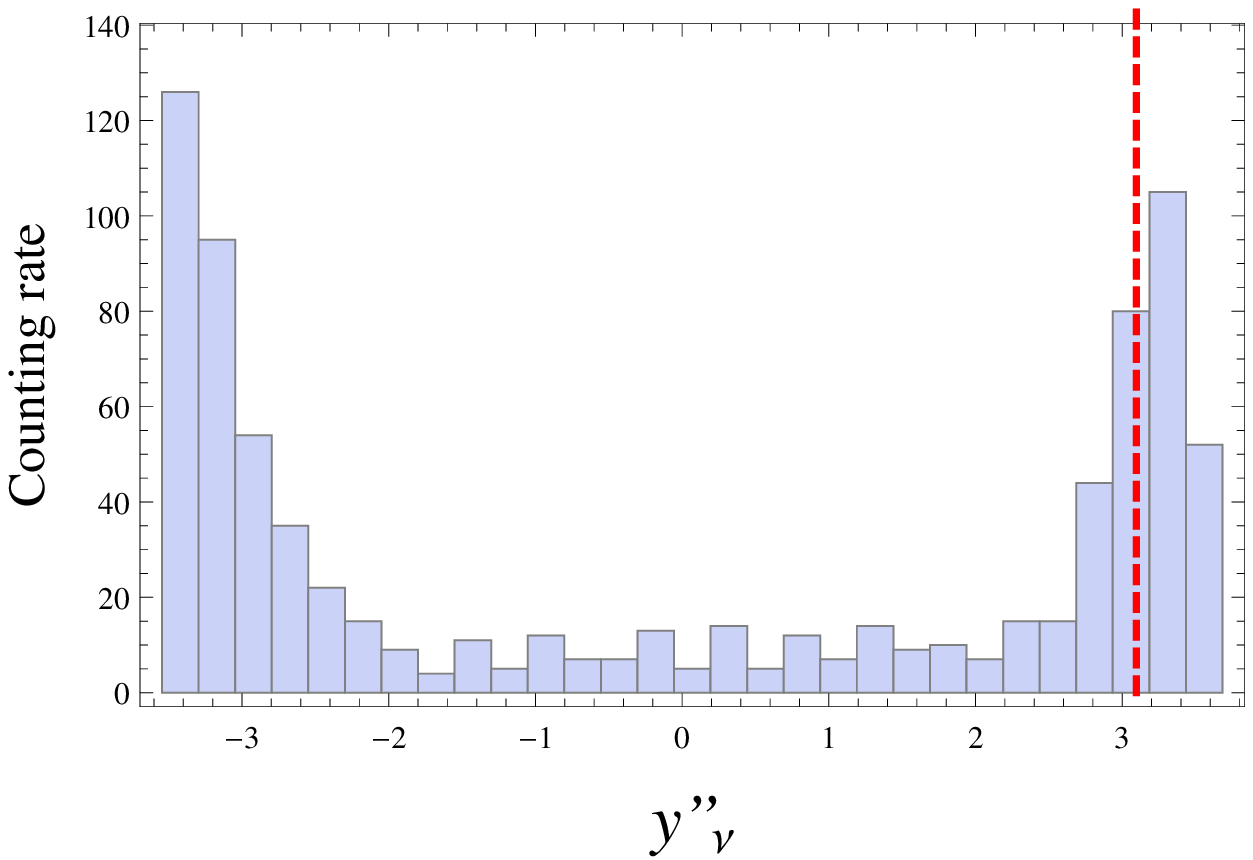} \\
		\includegraphics[scale=0.6]{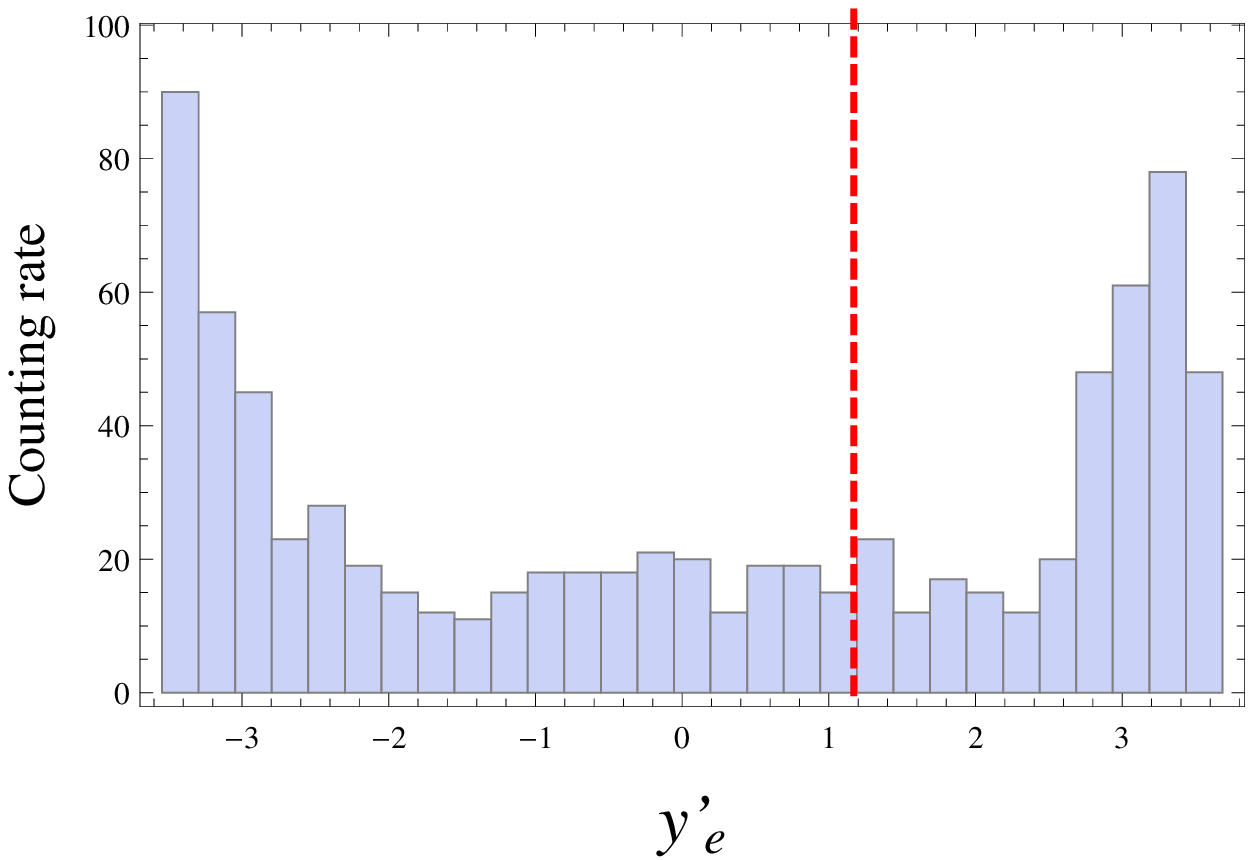} 
		\includegraphics[scale=0.6]{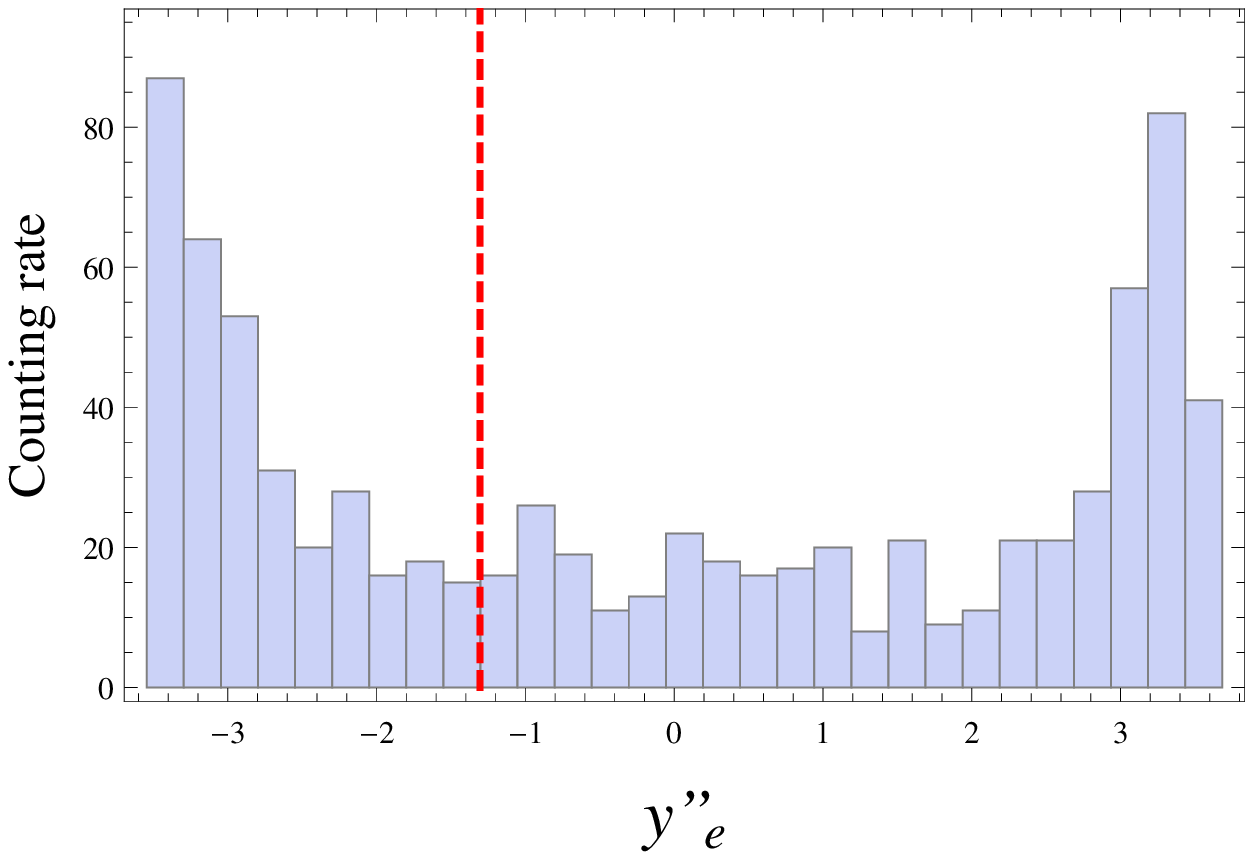} 
	\caption{\small{Counting rates for strong phase transition points as a function 
of the Yukawa couplings  $y'_\nu$ (top left), $y''_\nu$ (top right), $y'_e$ (bottom left), $y''_e$ (bottom right). The couplings of the benchmark point in eq. (\ref{BP}) are indicated as vertical lines.}}
	\label{fig:y}
\end{figure} 

 \begin{figure}[h]
	\centering
	    	\includegraphics[scale=0.6]{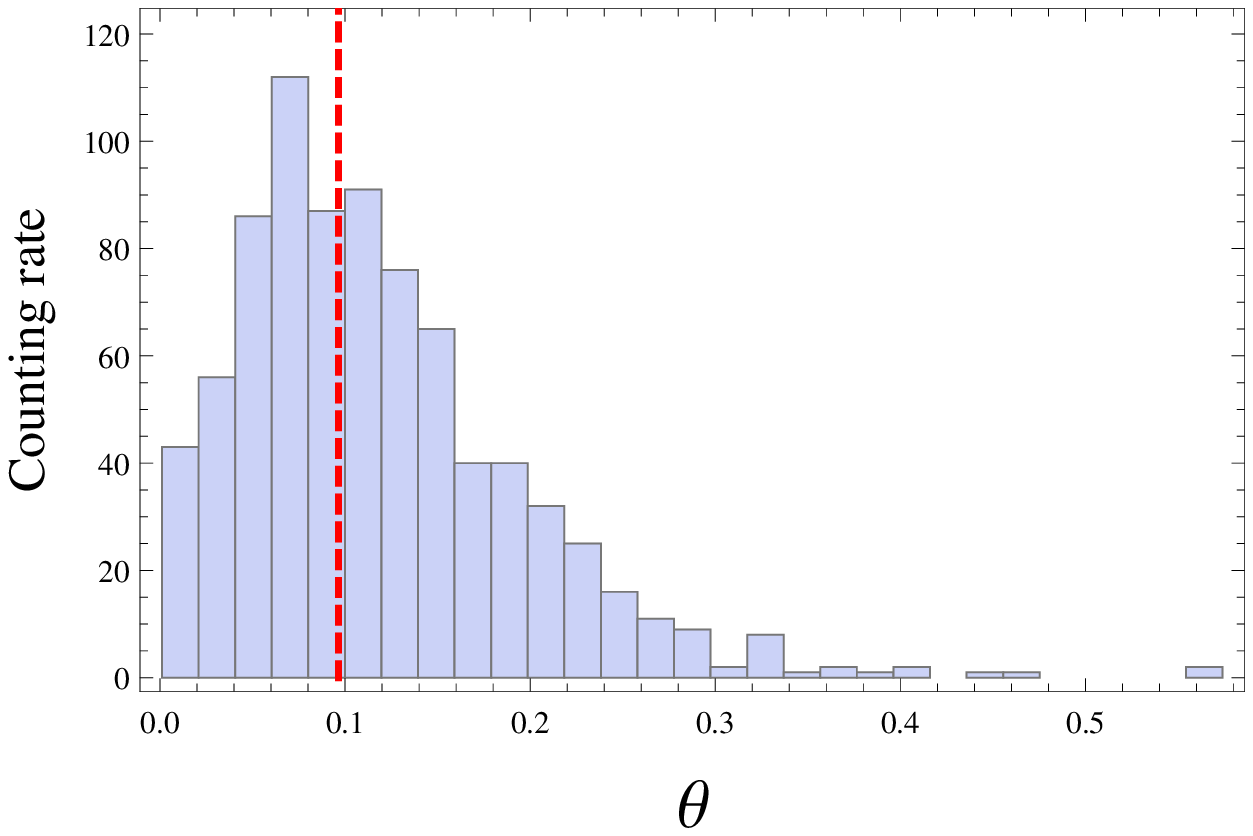} 
		\includegraphics[scale=0.6]{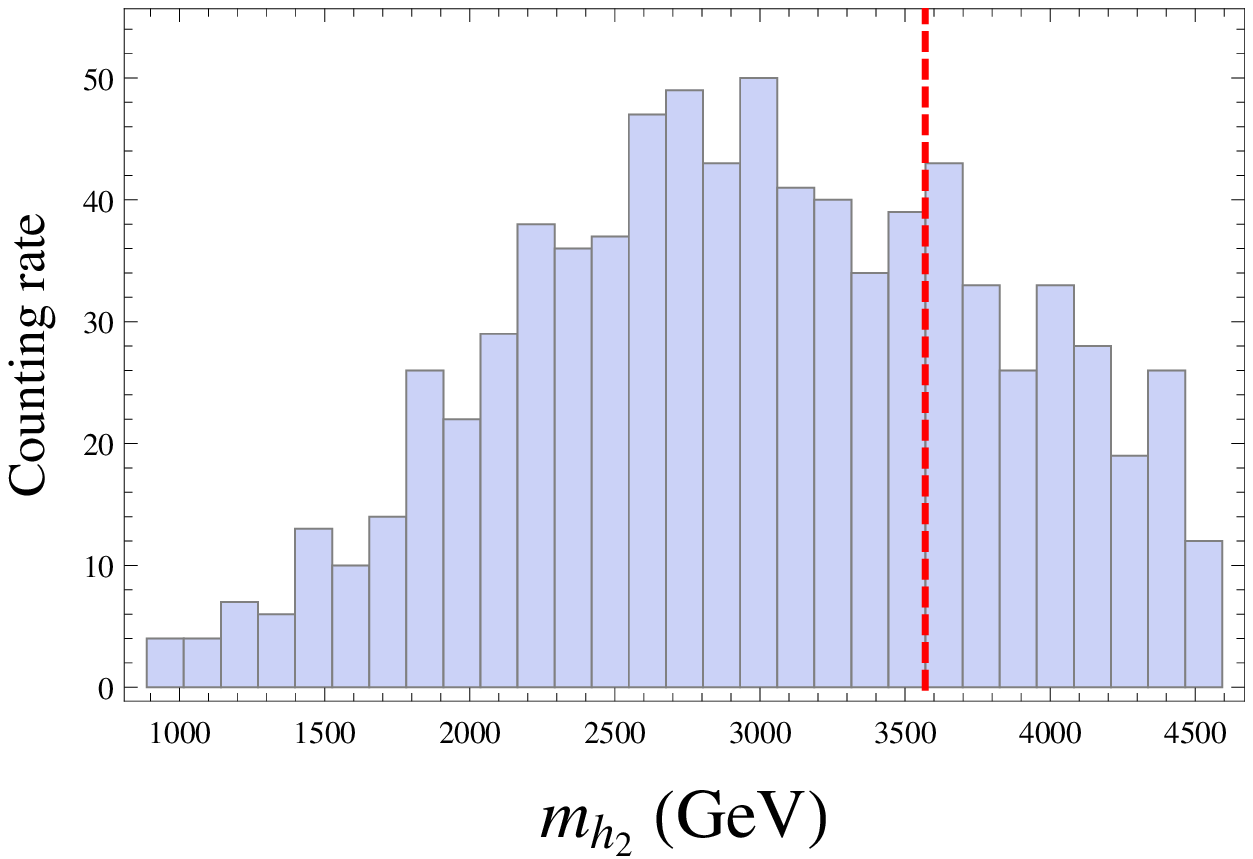} 
	\caption{\small{Counting rates for strong phase transition points as a function 
of the scalar mixing angle $\theta$ (left) and the mass of the heavy scalar $m_{h_2}$ (right). The values corresponding to the benchmark point in eq. (\ref{BP}) are indicated as  vertical lines.}}
	\label{fig:scpar}
\end{figure} 

 \begin{figure}[h]
	\centering
		\includegraphics[scale=.7]{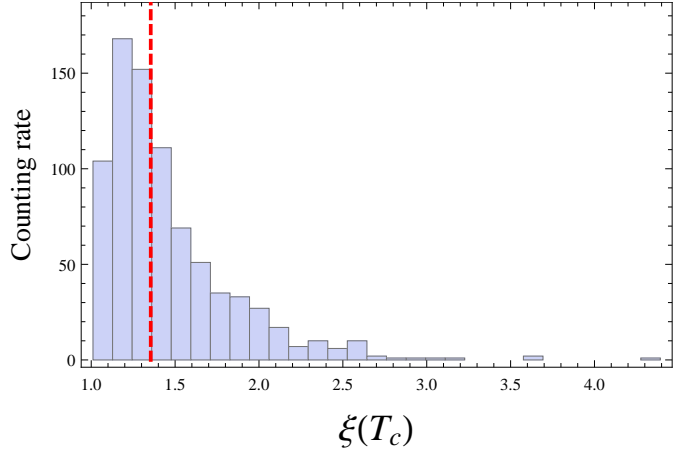} 
	\caption{\small{Frequency distribution of phase transition strength parameter $\xi(T_c)$ in the sample $\xi(T_c)>1$. The values corresponding to the benchmark point in eq. (\ref{BP}) are indicated as  vertical lines.}}
	\label{fig:xi}
\end{figure} 

\begin{figure}[h]
	\centering
	    	\includegraphics[scale=0.6]{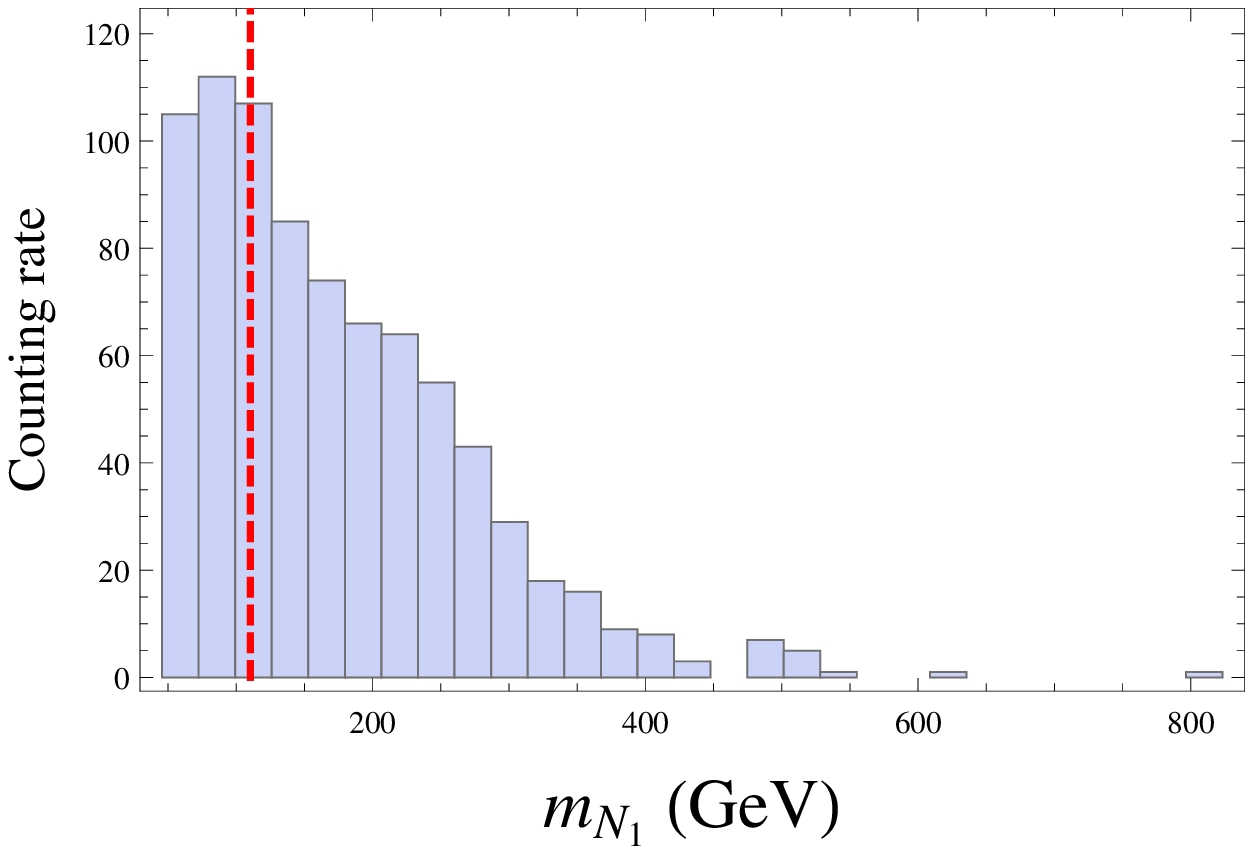} 
		\includegraphics[scale=0.6]{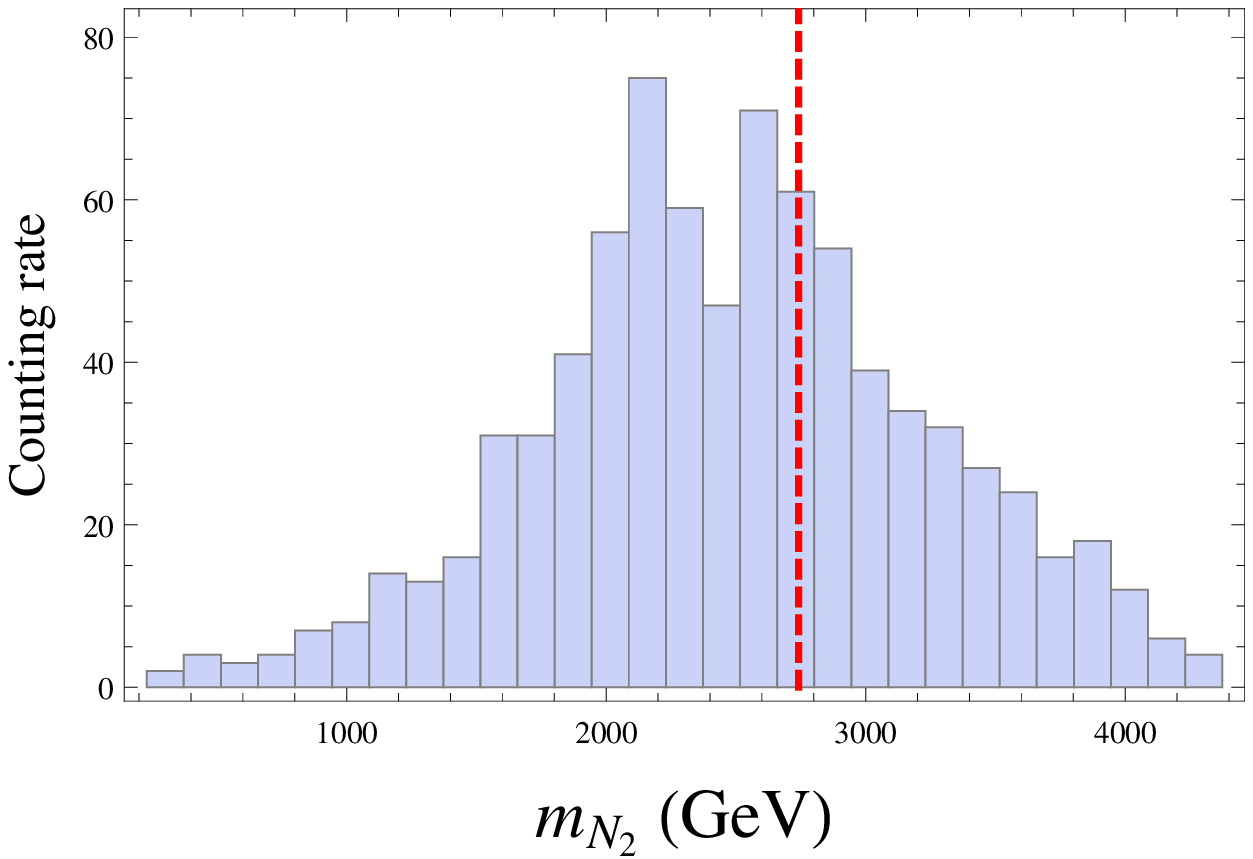} \\
		\includegraphics[scale=0.6]{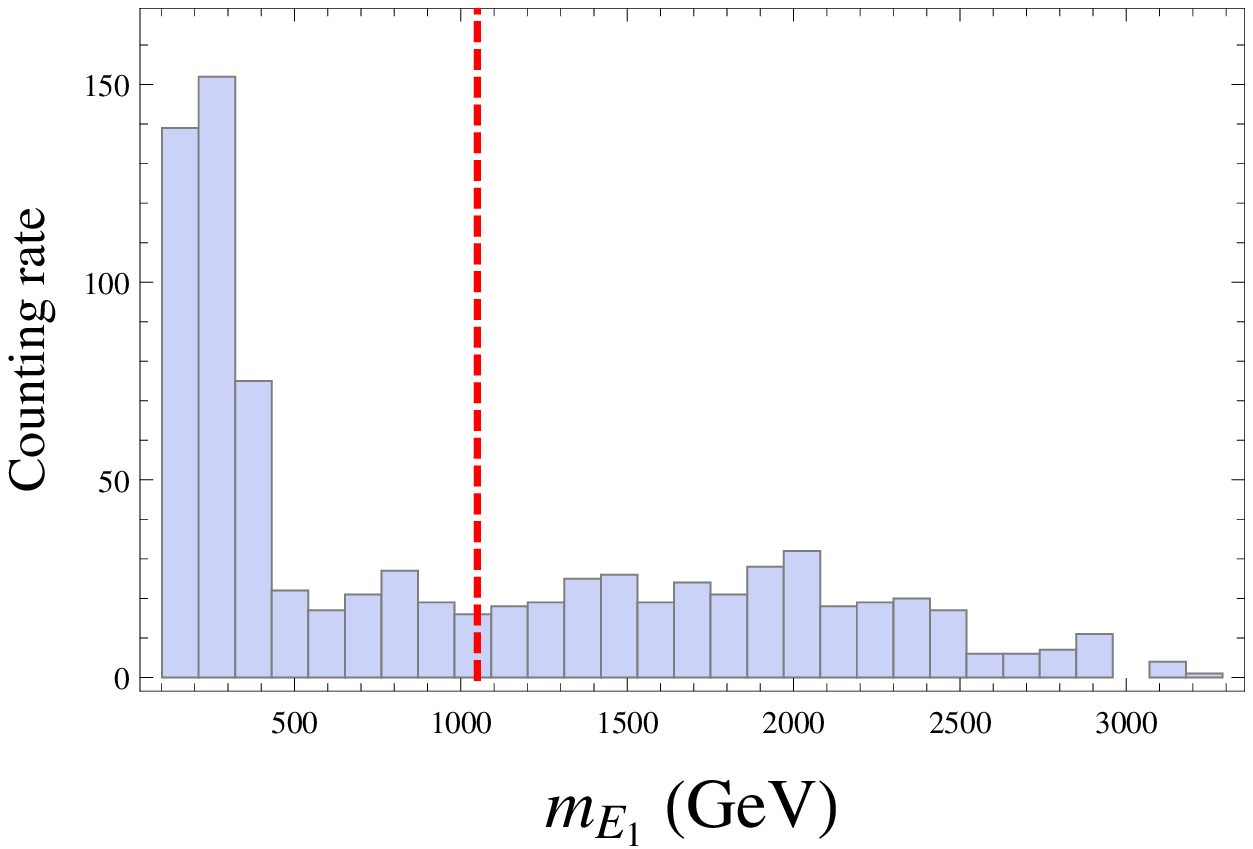} 
		\includegraphics[scale=0.6]{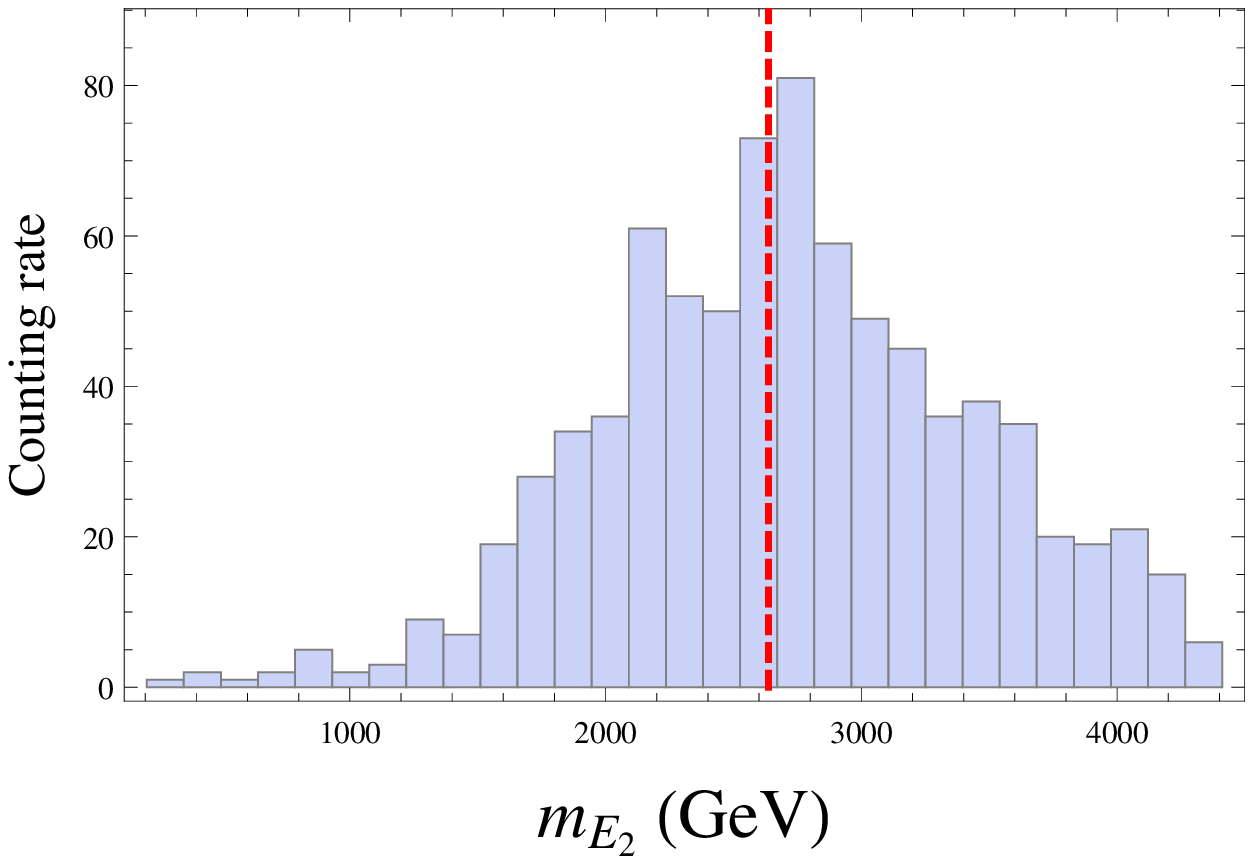}
\caption{\small{Counting rates for strong phase transition points as a function 
of the vector-like fermion masses  $m_{N_1}$ (top left), $m_{N_2}$ (top right), $m_{E_1}$ (bottom left), $m_{E_2}$ (bottom right). The values corresponding to the benchmark point in eq. (\ref{BP}) are indicated as  vertical lines.}}
\label{fig:masslep}
\end{figure}

From an initial set of $10^6$ randomly generated points, only $4.06\times 10^5$ survive the metastability test and can be considered as physical points.  Among them, a strong EWPT was found in only $809$ cases, that correspond to $0.2\%$ of the physical points. This implies that there is a narrow window for strong EWPT in the model.  

The obtained data sample of strong EWPT points is presented in histograms for each parameter of our model. As a reference, along with each histogram, we draw a vertical line representing the benchmark point
\begin{equation}\label{BP}
\begin{array} {c}
\lambda_1^{(BP)}=1.106\, ,\qquad \lambda_2^{(BP)}=2.184\, ,\qquad \lambda_3^{(BP)}=2.917\,,\\
c_\ell^{(BP)}= 2.183\, ,\qquad c_\nu^{(BP)}= -0.006\, ,\qquad c_e^{(BP)}=0.863\, ,\\
{y'_\nu}^{(BP)}= -3.417\, ,\qquad {y''_\nu}^{(BP)}=3.097\, , \qquad
{y'_e}^{(BP)}= 1.170 \, , \qquad {y''_e}^{(BP)} -1.305\, ,
\end{array}
\end{equation}
that has been chosen randomly from the dataset.   Figure \ref{fig:c} shows the frequency distributions of the $c_\nu$, $c_e$ and $c_\ell$ Yukawas in the parameter space corresponding to $\xi(T_c)>1$. Figure \ref{fig:y} shows the corresponding counting rates for the Yukawa couplings  $y'_\nu$, $y''_\nu$ , $y'_e$ and $y''_e$. The distributions of the two free parameters of the scalar sector are displayed in terms of the mixing angle $\theta$ and the mass of the heavy scalar $m_{h_2}$ in figure \ref{fig:scpar}.  Finally, the frequency counts of $\xi(T_c)$ values are plotted in figure \ref{fig:xi}, where it can be appreciated that only mild transitions can be obtained within this setup.

From figure \ref{fig:c}, we observe that the preferred values for  $c_\nu$  tend to accumulate close to zero. This effect reduces the total mass of the new neutrinos.  A similar behavior is seen for $c_\ell$, but in this case, there are two additional peaks around $c_\ell=\pm 2$.  On the other hand, figure \ref{fig:y} shows that a strong EWPT is favored whenever the $y$-type Yukawas (prominently those that belong to the neutral leptons) take values near the perturbative limit. We can interpret this phenomenon as follows: significant contributions of the vector-like leptons are required in order to obtain a strong transition, and therefore large Yukawa couplings are expected for this model. However, there is a tension between large Yukawas and large lepton masses, as a substantial mass for any field enters quickly in the decoupling limit. Hence, a reasonably low mass for the new leptons, with sizable Yukawa contributions, can be achieved by setting a $c$-type Yukawa (which couples to the lepton number breaking scale $u$) with natural values $(0,\pm\sqrt{4\pi}/2)$, and pushing the $y$-type Yukawas (coupled to the electroweak scale $v$) all the way up to the perturbative limit $\pm\sqrt{4\pi}$. The consistency of the model under the presence of such large Yukawa couplings, and their contribution to the dark matter relic density deserve further study, to be presented in a future work.

In the scalar sector, figure \ref{fig:scpar} shows that a strong EWPT prefers a mixing angle peaked around $\theta\sim 0.08$, and that there is a broad distribution for the heavy scalar masses. Nevertheless, the net effect of the scalar sector in the phase transition is overshadowed by the contribution of the vector-like fermions. 

To end this section, we present in figure \ref{fig:masslep} the distributions for the masses of the vector-like leptons in the sample of strong EWPT points. In this figure, it can be noticed that the masses of the lightest neutral and charged fermions required for a strong EWPT are prone to accumulate at their lower experimental bound. 

\section{Conclusions}\label{secCon}

In this work, the EWPT in the minimal $SU(3)_c\times SU(2)_W\times U(1)_Y\times U(1)_L$ extension of the SM was studied. Anomaly cancellation requires the introduction of at least one family of vector-like leptons, that can contribute to the phase transition. The model was found to be able to provide a strong transition for a very limited number of cases. The electroweak Yukawa couplings of the new vector-like fermions tend to be close to their perturbative limit. This effect puts serious restrictions to the viability of the model as a candidate to explain the baryon asymmetry of the universe. In this context, the restrained parameter space in which electroweak baryogenesis is possible, can be significantly reduced in the near future, e.g. by the determination of stricter bounds on the mass of an exotic charged lepton, which in our analysis tends to be light. Yet, the model provides the needed equilibrium departure required by the Sakharov conditions in its present form, and at the same time yields a contribution for the DM content of the universe.  A possible enhancement of the EWPT strength in this scheme is the inclusion of more families of vector-like leptons, which can alleviate the tension between the low masses for the new fermions and  the large Yukawa couplings needed for a strong transition.

\appendix
\section{Field dependent squared masses}\label{apA}

\begin{itemize}

\item Scalars: $n_{h_{1,2}}=1$,
\begin{equation}
\begin{split}
&m_{h_{1,2}}^2(\phi_1,\phi_2,T)=\frac{1}{2}\left(A_h\mp B_h\right)\, ,\\
&A_h= \lambda_1 \left(3 \phi_1^2-v^2\right)+\lambda_2 \left(3 \phi_2^2-u^2\right)+\frac{\lambda_3}{2}  \left(\phi_1^2-v^2+\phi_2^2-u^2\right)\\
&\qquad+\kappa_1(T)+\kappa_2(T)\, ,\\
&B_h=\Big\{4\lambda_3^2\phi_1^2\phi_2^2+\Big[\lambda_2 \left(3 \phi_2^2-u^2\right)-\lambda_1 \left(3 \phi_1^2-v^2\right)\\
&\qquad+\frac{\lambda_3}{2}  \left(\phi_1^2-v^2-\phi_2^2+u^2\right)+\kappa_2(T)-\kappa_1(T)\Big]^2\Big\}^{1/2}\, ,\\
&\kappa_1=\frac{T^2}{48}\left[9g_W^2+3g_Y^2+12y_t^2+24\lambda_1+4\lambda_3+4\left({y_\nu'}^2+{y_\nu''}^2+{y_e'}^2+{y_e}''^2\right)\right]\, ,\\
&\kappa_2=\frac{T^2 }{12} \left(2 c_\ell^2+c_\nu^2+c_e^2+27 g_L^2+4 \lambda_2+2\lambda_3\right)\, .
\end{split}
\end{equation}

\item Goldstone bosons: $n_{G^\pm}=2$, $n_{G^0_{1,2}}=1$,
\begin{equation}
\begin{split}
m_{G^\pm,G_1^0}^2(\phi_1,\phi_2,T)=&\lambda_1 \left(\phi_1^2-v^2\right)+\frac{\lambda_3}{2} \left(\phi_2^2-u^2\right)+\kappa_1(T)\, ,\\
m_{G_2^0}^2(\phi_1,\phi_2,T)=&   \lambda_2 \left(\phi_2^2-u^2\right)+\frac{\lambda_3}{2} \left(\phi_1^2-v^2\right)+\kappa_2(T)\, .
\end{split}   
\end{equation}

\item New gauge boson: $n_{Z_L}=3$,
\begin{equation}
m_{Z_L}^2(\phi_1,\phi_2)=9g_L^2\phi_2^2,
\end{equation} 

\item Neutral vector-like leptons: $n_{N_{1,2}}=-4$,
\begin{equation}
\begin{split}
&m_{N_1,N_2}^2(\phi_1,\phi_2,T)=\frac{1}{4}\left(A_N\mp B_N\right)\, ,\\
&A_N=\phi_1^2 \left(y_\nu'^2+y_\nu''^2\right)+\phi_2^2 \left(c_\ell^2+c_\nu^2\right)\, ,\\
&B_N=\Big\{\left[\phi_1^2 (y_\nu'+y_\nu'')^2+\phi_2^2 (c_\ell-c_\nu)^2\right] \left[\phi_1^2 (y_\nu'-y_\nu'')^2+\phi_2^2
   (c_\ell+c_\nu)^2\right]\Big\}^{1/2}\,  .
\end{split}   
\end{equation}

\item Charged vector-like leptons: $n_{E_{1,2}}=-4$
\begin{equation}
\begin{split}
&m_{E_1,E_2}^2(\phi_1,\phi_2,T)=\frac{1}{4}\left(A_E\mp B_E\right)\, ,\\
&A_E=\phi_1^2 \left(y_e'^2+y_e''^2\right)+\phi_2^2 \left(c_\ell^2+c_e^2\right)\, ,\\
&B_E=\Big\{\left[\phi_1^2 (y_e'+y_e'')^2+\phi_2^2 (c_\ell-c_e)^2\right] \left[\phi_1^2 (y_e'-y_e'')^2+\phi_2^2
   (c_\ell+c_e)^2\right]\Big\}^{1/2}\, .
\end{split}   
\end{equation}

\end{itemize}

\acknowledgments{This work was supported in part by CONACYT and SNI (Mexico).  CAV-A acknowledges the Mainz Institute for Theoretical Physics (MITP) for hospitality and support during the revision stage of this work.}

\end{document}